\documentclass[preprint2,times,tighten]{aastex6}
\usepackage{amsmath,amstext}
\usepackage[figure,figure*]{hypcap}
\usepackage{newtxmath} %use times font for math

\journalinfo{The Astrophysical Journal Letters}

\newcommand{\Msun}{\ensuremath{M_\odot}}

\newcommand{\kmps}{\ensuremath{\mathrm{km~s^{-1}}}}

\newcommand{\Ledd}{\ensuremath{L_\mathrm{Edd}}}
\newcommand{\Hbeta}{\ensuremath{\mathrm{H{\beta}}}}

\newcommand{\Lhb}{\ensuremath{L(\Hbeta)}}

\shorttitle{STACs by BH disk winds}
\shortauthors{Moriya et al.}

\begin{document}

\title{
Superluminous transients at AGN centers from interaction between black-hole disk winds and broad-line region clouds
}

\author{Takashi J. Moriya\altaffilmark{1},
Masaomi Tanaka\altaffilmark{1},
Tomoki Morokuma\altaffilmark{2},
Ken Ohsuga\altaffilmark{1}
} 
\email{takashi.moriya@nao.ac.jp}
\altaffiltext{1}{Division of Theoretical Astronomy, National Astronomical Observatory of Japan, National Institutes of Natural Sciences, 2-21-1 Osawa, Mitaka, Tokyo 181-8588, Japan}
\altaffiltext{2}{Institute of Astronomy, Graduate School of Science, The University of Tokyo, 2-21-1 Osawa, Mitaka, Tokyo 181-0015, Japan}

\begin{abstract}
 We propose that superluminous transients that appear at central regions of
 active galactic nuclei (AGNs) such as CSS100217:102913+404220
 (CSS100217) and PS16dtm, which reach near or super-Eddington luminosities of the central black holes, are powered by the interaction between accretion disk winds and clouds in broad-line regions (BLRs)
 surrounding them.
 If the disk luminosity temporary increases
 by, e.g., limit-cycle oscillations, leading to 
 a powerful radiatively driven wind,
 strong shock waves propagate in the BLR.
 Because the dense clouds in the AGN BLRs typically have similar densities to those found in Type~IIn supernovae, strong radiative shocks emerge and efficiently convert the ejecta kinetic energy to radiation. As a result, transients similar to Type~IIn supernovae can be observed at AGN central regions. Since a typical black-hole disk wind velocity is $\simeq 0.1c$ where $c$ is the speed of light, the ejecta kinetic energy is expected to be $\simeq 10^{52}~\mathrm{erg}$ when $\simeq 1~\Msun$ is ejected. This kinetic energy is transformed to radiation energy in a timescale for the wind to sweep up a similar mass to itself in the BLR, which is a few hundred days. Therefore, both luminosities ($\sim 10^{44}~\mathrm{erg~s^{-1}}$) and timescales ($\sim 100$~days) of the superluminous transients from AGN central regions match to those expected in our interaction model. If CSS100217 and PS16dtm are related to the AGN activities triggered by limit-cycle oscillations, they become bright again in coming years or decades.
\end{abstract}


\section{Introduction}
Our knowledge of the transient Universe is expanding thanks to recent large untargeted transient surveys like Palomer Transient Factory \citep[e.g.,][]{law2009ptf}, Pan-STARRS1 \citep[e.g.,][]{kaiser2010ps1}, Catalina Real-Time Transient Survey \citep{drake2009css}, Kiso Supernova Survey \citep[e.g.,][]{morokuma2014kiss}, and All-Sky Automated Survey for Supernovae \citep[e.g.,][]{shappee2014asassn}. An example of new discoveries brought by such surveys is the discovery of superluminous supernovae (SLSNe, \citealt{quimby2011slsn}). SLSNe are supernovae (SNe) more than $\sim 10$ times brighter than canonical SNe and their peak luminosities exceed $\sim 10^{44}~\mathrm{erg~s^{-1}}$ or $-21~\mathrm{mag}$ in optical bands (see \citealt{howell2017slsnreview} for a recent review). 

While many SLSNe appear in the outskirts of their host galaxies and their progenitors are likely massive stars, some of them, especially the brightest ones, are often found at the central regions of their host galaxies where telescopes' limited resolutions make it impossible to judge whether they are off-center or not. For example, the brightest known SLSN candidate ASASSN-15lh \citep{dong2016asassn15lh} is from the center of the host galaxy and it may actually be a tidal disruption event (TDE) rather than a SLSN \citep{leloudas2016asassn15lh}. Another luminous transient called ``Dougie'' also appeared at the center of its host galaxy and is interpreted as a TDE \citep{vinko2015dougie}.

\citet{drake2011css} reported a discovery of a SLSN candidate at the central region of an active galactic nucleus (AGN). The SLSN candidate is named CSS100217:102913+404220 (CSS100217 hereafter) and it reached the peak optical magnitude of around $-23~\mathrm{mag}$, corresponding $\sim 5\times 10^{44}~\mathrm{erg~s^{-1}}$. CSS100217 kept its brightness within one magnitude from the peak for $\sim 100~\mathrm{days}$ and its total radiated energy was $1.3\times 10^{52}~\mathrm{erg}$. Although CSS100217 appeared at the AGN central region, \citet{drake2011css} suggest that CSS100217 is not associated with the host AGN activity because the magnification amplitude and the transient timescale of CSS100217 are not expected from typical AGN activities. Because the host AGN is a narrow-lined Seyfert 1 (NLS1) galaxy where star formation activities may be enhanced near the central black hole (BH), \citet{drake2011css} suggested that CSS100217 is a genuine SLSN from a massive star that happened to appear near the AGN central region due to the enhanced star formation there. Indeed, CSS100217 had spectra which are similar to those of Type~IIn SLSNe. Type~IIn SLSNe are believed to be explosions of very massive stars with circumstellar media of $\sim 10~\Msun$ \citep[e.g.,][]{moriya2013sn2006gy}.

\citet{blanchard2017ps16dtm} recently reported another transient PS16dtm that had Type~IIn-like spectra and appeared at the central region of a NLS1 galaxy. Interestingly, both CSS100217 and PS16dtm reach near or above the Eddington luminosities of the central BHs of the hosting AGN \citep{blanchard2017ps16dtm}. \citet{blanchard2017ps16dtm} suggest that CSS100217 and PS16dtm are TDEs instead of massive star explosions. However, 
the fact that similar transients preferentially appear in AGN central regions 
with similar properties motivated us to consider a possibility that these transients are 
related to AGN activities. In the rest of this paper, we call such luminous transients from AGN central regions as ``Superluminous Transients from AGN Central regions'' or ``STACs''. We suggest that STACs are caused by the interaction between a BH accretion disk wind and the clouds in the broad-line region (BLR) surrounding the BH. 

We assume the standard cosmology with $H_0 = 70~\mathrm{km~s^{-1}~Mpc^{-1}}$, $\Omega_M =0.3$, and $\Omega_\Lambda =0.7$.

\section{AGN properties}\label{sec:agnproperties}
We first summarize properties of BLRs in AGNs. BLRs locate at the immediate vicinity of the central BHs in AGNs. If there is an accretion disk wind from the central BH, the wind propagates in the BLR.

The radii of BLRs containing hydrogen ($R_\mathrm{BLR}$) can be empirically estimated by their luminosities at 5100~\AA\ ($\lambda L_{5100}$), i.e., \citep[e.g.,][]{bentz2013agn}
\begin{equation}
R_\mathrm{BLR} \simeq 0.03 \left(\frac{\lambda L_{5100}}{10^{44}~\mathrm{erg~s^{-1}}}\right)^{0.53}\ \mathrm{pc}.
\label{eq:Rblr}
\end{equation}
The mass of ionizing materials in BLRs ($M_\mathrm{BLR}$) can be estimated with the \Hbeta\ luminosity, \Lhb\ \citep{osterbrock2006agn}. The \Hbeta\ luminosity can be expressed as
\begin{equation}
\Lhb \simeq n_e n_p \alpha_\mathrm{H\beta}^\mathrm{eff} h \nu_\mathrm{H\beta}V\varepsilon,
\label{eq:Lhb}
\end{equation}
where $n_e$ is the electron number density, $n_p$ is the proton number density, $\alpha_\mathrm{H\beta}^\mathrm{eff}$ is the effective recombination efficiency, $h$ is the Planck constant, $\nu_\Hbeta$ is the \Hbeta\ frequency, $V$ is the volume of the BLR, and $\varepsilon$ is the filling factor. The filling factor represents the volume fraction of the high density clouds in the BLRs (Fig.~\ref{fig:stacpicture}). Using Equation~(\ref{eq:Lhb}), we can obtain the BLR mass $M_\mathrm{BLR}$ as
\begin{eqnarray}
M_\mathrm{BLR} &\simeq& (n_p m_p + n_\mathrm{He} m_\mathrm{He})V\varepsilon, \\
&\simeq& 0.9 \left(\frac{\Lhb}{10^{41}~\mathrm{erg~s^{-1}}} \right) \left(\frac{n_e}{10^9~\mathrm{cm^{-3}}} \right)^{-1} \Msun, \label{eq:massblr} \\
\nonumber
\end{eqnarray}
where $m_p$ is the proton mass, $n_\mathrm{He}$ is the helium number density, $m_\mathrm{He}$ is the helium mass. We assume $n_\mathrm{He}\simeq 0.1 n_p$ in deriving Equation~(\ref{eq:massblr}). This assumption holds when the half of the helium is He$^+$ and the other half is He$^{++}$ in a solar-metallicity gas. $n_e$ is roughly $\simeq 10^9~\mathrm{cm^{-3}}$ in AGNs \citep{osterbrock2006agn}.
With the density, mass, and radius of BLRs, we can estimate the filling factor as
\begin{equation}
\varepsilon \simeq 3\times 10^{-4}\left(\frac{\lambda L_{5100}}{10^{44}~\mathrm{erg~s^{-1}}}\right)^{-1.6}\left(\frac{\Lhb}{10^{41}~\mathrm{erg~s^{-1}}} \right) \left(\frac{n_e}{10^9~\mathrm{cm^{-3}}} \right)^{-2}.
\label{eq:fillingfactor}
\end{equation}
The covering factor $f_c$ of the BLR clouds, i.e., the fraction of sight-lines covered by the BLR clouds, is determined by $\varepsilon$ and the cloud size. It is typically estimated to be $f_c \simeq 0.4$ \citep[e.g.,][]{dunn2007blrcoveringfac,gaskell2008blrcoveringfac}.

With the virial theorem and the relation between $R_\mathrm{BLR}$ and the AGN continuum luminosity (Eq.~\ref{eq:Rblr}), we can obtain the following relation to estimate the BH mass at an AGN center \citep{mclure2004agn},
\begin{equation}
M_\mathrm{BH} \simeq 6\times 10^6 \left(\frac{\lambda L_{5100}}{10^{44}~\mathrm{erg~s^{-1}}} \right)^{0.53}\left(\frac{\mathrm{FWHM}(\Hbeta)}{1000~\kmps}\right)^2\ \Msun,
\label{eq:massbh}
\end{equation}
where $\mathrm{FWHM}(\Hbeta)$ is the full-width half-maximum (FWHM) velocity of the emission line of \Hbeta.

Finally, the Eddington luminosities of BHs, at which the gravitational and radiation pressures are balanced, are 
\begin{equation}
\Ledd = \frac{4\pi G c M_\mathrm{BH}}{\kappa} \simeq 10^{38}\frac{M_\mathrm{BH}}{\Msun}~\mathrm{erg~s^{-1}},
\label{eq:Ledd}
\end{equation}
where $G$ is the gravitational constant and $\kappa\simeq 0.34~\mathrm{cm^2~g^{-1}}$ is the electron-scattering opacity.

\section{The BH disk wind model}\label{sec:stacsbyagn}

\subsection{General picture}\label{sec:generalpicture}
The basic idea of our BH disk wind model is to make STACs bright through the interaction between the BH disk wind and the clouds in the BLR surrounding the central BH. Our model is mainly composed of two phenomena: (i) the mass ejection from the BH disk and (ii) the interaction between the ejected mass and the BLR clouds. 

At first, at the stage (i), a strong wind from the central BH accretion disk with the mass $M_\mathrm{ej}^\mathrm{BH}$ and the velocity $v_\mathrm{ej}^\mathrm{BH}$ is launched. Our model is independent of the wind launching mechanism, but we suggest that the BH disk wind is triggered by 
the limit-cycle oscillation which is induced by the thermal-viscous instability when the mass accretes onto the BH with the near-Eddington rate (Section~\ref{sec:bhmassejection}). This is because the observed luminosities of STACs are near- or super-Eddington. In addition to the sudden increase of the disk luminosity, the limit-cycle oscillation can make the mass ejection rate temporarily more than 10 times larger than the Eddington rate \citep[e.g.,][]{ohsuga2006limitcycle}. A typical velocity of the BH wind is $v_\mathrm{ej}\simeq 0.1c$ where $c$ is the speed of light. The timescale of the activity is $\sim10-100$~sec for stellar mass BHs, and is considered to be $\sim 10-1000$~days for $M_\mathrm{BH}\sim 10^6-10^8~\Msun$ (Section~\ref{sec:bhmassejection}). For the central BH mass of $10^7~\Msun$, 
if the mass ejection with the Eddington rate ($L_\mathrm{Edd} \simeq 10^{45}~\mathrm{erg~s^{-1}}$) occurs for 100~days, the total kinetic energy output is $E_\mathrm{BH} \sim 10^{52}~\mathrm{erg}$.
This is comparable to the typical total radiated energy of STACs. In our model, this powerful outflow powers STACs. For the velocity of $v_\mathrm{ej}^\mathrm{BH} \simeq 0.1c$, the ejected mass needs to be $M_\mathrm{ej}^\mathrm{BH} \simeq 1~\Msun$ ($E_\mathrm{kin}^\mathrm{BH}=M_\mathrm{ej}^\mathrm{BH}v_\mathrm{ej}^{\mathrm{BH}\ 2}/2$).

Although the disk luminosity can exceed the Eddington luminosity during the limit-cycle oscillation, the disk wind will immediately obscure the luminous disk. In the above example of $M_\mathrm{BH}=10^{7}~\Msun$, $\sim 10^{-2}~\Msun$ is ejected in a day. With the velocity of $\simeq 0.1c$, it reaches $\sim 3\times 10^{14}~\mathrm{cm}$. If we assume that the spherically-symmetric fully-ionizing gas flows as the BH disk wind, we can apply the same approximation to estimate its opacities as in \citet{roth2016tdeopacity}. Assuming that the inner radius of the ejecta is $\sim 10^{13}$~cm, the electron scattering optical depth of the ejecta at one day is $\tau_\mathrm{es}\simeq 180$. Because the absorption optical depth ($\tau_\mathrm{ab}$) is $\sim 10^{-4}$ times smaller than the scattering optical depth \citep{roth2016tdeopacity}, the effective optical depth in the BH disk wind is $\tau_\mathrm{eff}=\sqrt{\tau_\mathrm{sc}\tau_\mathrm{ab}}\sim 2$. Therefore, at 1~day after the mass ejection, the BH disk wind can already obscure the central X-ray and ultraviolet source.
As long as the mass ejection continues, the central source is kept obscured via the wind with $\tau_\mathrm{eff}$ of the order of unity and the accretion disk does not directly contribute to the luminosities of STACs in our model.

\begin{figure}
 \begin{center}
  \includegraphics[width=0.9\columnwidth]{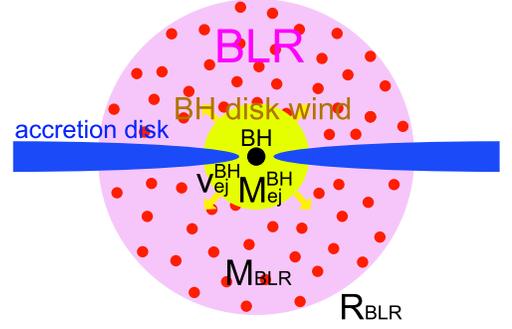}  
 \end{center}
\caption{
Schematic picture of our BH disk wind model. If the BH accretion disk ejects a wind with $M_\mathrm{ej}^\mathrm{BH}$ and $v_\mathrm{ej}^\mathrm{BH}$, it is decelerated by the dense clouds in the BLRs (red dots) and its kinetic energy is efficiently converted to radiation.
}\label{fig:stacpicture}
\end{figure}

Now we consider the second phenomenon, i.e., (ii) the interaction between the BH disk wind and the BRL clouds. We illustrate this phase in Fig.~\ref{fig:stacpicture}. When $n_e\simeq 10^9~\mathrm{cm^{-3}}$ in the BLR clouds, the corresponding cloud density is $\simeq 2\times 10^{-15}~\mathrm{g~cm^{-3}}$. This density is as high as those found in Type~IIn SNe \citep[e.g.,][]{moriya2014iinmasslosshist}. Therefore, similar physical processes to those in Type~IIn SNe are presumed to occur in these clouds when BH disk wind collides to them. SN ejecta are replaced with BH disk winds and dense circumstellar media are replaced with BLR clouds. Strong radiative shock waves are created between the BH disk wind and the BLR clouds. High-energy photons from the shock are immediately absorbed by the clouds due to their high density and heat up the clouds. Most of radiation is emitted in near-ultraviolet to optical wavelengths and the shocks also make narrow lines as in Type~IIn SNe. The efficient conversion from the kinetic energy to radiation at the shocks makes the BH disk wind bright and results in STACs.

The emission timescale from the interaction ($t_\mathrm{em}$) is determined by the timescale for the BH wind to lose its kinetic energy.
The deceleration timescale corresponds to the timescale for the shock to sweep the comparable mass to the BH disk wind. Keeping in mind that only $f_c M_\mathrm{ej}$ interacts with the BLR clouds, we obtain the following emission timescale; \begin{equation}
t_\mathrm{em}\simeq 240 \left(\frac{f_c M_\mathrm{ej}^\mathrm{BH}}{\Msun}\right)^{\frac{1}{3}}\left(\frac{v_\mathrm{ej}^{\mathrm{BH}}}{0.1c}\right)^{-1}\left(\frac{\varepsilon}{10^{-3}}\right)^{-\frac{1}{3}}\left(\frac{n_e}{10^9~\mathrm{cm^{-3}}}\right)^{-\frac{1}{3}}~\mathrm{days}.
\label{eq:tem}
\end{equation}
The light-crossing time of the shock is $\sim 10$ times smaller than $t_\mathrm{em}$ and it does not affect STAC timescales.

We briefly summarize the general picture. First, the BH disk wind is launched by a disk instability like limit-cycle oscillations. Although the disk will be bright due to the instability, the BH disk wind becomes optically thick and the central disk is obscured as long as the mass ejection continues. The mass ejection typically continues for about 100~days or less. The BH disk wind becomes bright enough to be observed as STACs due to its collision to the surrounding clouds in the BLRs which efficiently converts kinetic energy to radiation. The emission timescale is determined by the deceleration timescale which is typically more than 100~days.

\subsection{CSS100217}\label{sec:css100217}
We apply our BH disk wind model to CSS100217. We start by looking into the AGN properties hosting CSS100217.

The spectrum of the galaxy hosting CSS100217 obtained by Sloan Digital Sky Survey (SDSS) suggests that it is a Seyfert and therefore an AGN \citep{abazajian2009sdss7}. The AGN has the $g$-band magnitude of 17.9~mag and its redshift is 0.147 \citep{drake2011css}. Thus, $\lambda L_{5100}\simeq 9\times 10^{43}~\mathrm{erg~s^{-1}}$. Its bolometric luminosity $L_\mathrm{bol}$ is $\simeq 7\times 10^{44}~\mathrm{erg~s^{-1}}$, if we apply the bolometric correction of $L_\mathrm{bol}\simeq 8.1\lambda L_{5100}$ \citep{runnoe2012agbbolcorr}. The broad component of its H$\beta$ emission has 2900~\kmps\ (FWHM) with the luminosity of $\Lhb\simeq 4\times 10^{41}~\mathrm{erg~s^{-1}}$ \citep{drake2011css}. Adopting the equations in Section~\ref{sec:agnproperties}, we estimate $R_\mathrm{BLR}\simeq 0.03~\mathrm{pc}$, $M_\mathrm{BLR}\simeq 4~\Msun$, $\varepsilon\simeq 10^{-3}$, and $M_\mathrm{BH}\simeq 5\times 10^7~\Msun$. Therefore, the Eddington luminosity is $\simeq 6\times 10^{45}~\mathrm{erg~s^{-1}}$. Thus, the Eddington ratio $\Gamma=L_\mathrm{bol}/\Ledd$ is about 0.1 at the central BH.

The integrated radiation energy emitted by CSS100217 is $\simeq 1.3\times 10^{52}~\mathrm{erg}$ \citep{drake2011css}. Assuming that this kinetic energy originates from the ejecta from the BH disk wind and only $f_c M_\mathrm{ej}$ $(f_c\simeq 0.4)$ interacts with the BLR clouds, $E_\mathrm{kin}^\mathrm{BH}\simeq 3\times 10^{52}~\mathrm{erg}$ is required. With the typical BH wind velocity of $v_\mathrm{ej}^\mathrm{BH}\simeq 0.1c$ \citep[e.g.,][]{hashizume2015bhwind}, the required total ejecta mass is estimated as $M_\mathrm{ej}^\mathrm{BH}\simeq 4~\Msun$.

We find that $f_cM_\mathrm{ej}^\mathrm{BH}\simeq 2~\Msun$ interacts with the BLR clouds. Because $M_\mathrm{BLR}\simeq 4~\Msun\gtrsim f_cM_\mathrm{ej}^\mathrm{BH}\simeq 2~\Msun$, the BLR region contains enough amount of matter to decelerate the BH disk wind. The emission timescale $t_\mathrm{em}$ is $t_\mathrm{em}\simeq 280~\mathrm{days}$. 
This timescale roughly matches that observed in CSS100217. Overall, the properties of CSS100217 can be explained by our BH disk wind model and CSS100217 can be powered by the interaction between the BH disk wind and the BLR clouds.

\subsection{PS16dtm}\label{sec:ps16dtm}
The host AGN of PS16dtm has $\lambda L_{5100}\simeq 10^{43}~\mathrm{erg~s^{-1}}$, $\Lhb\simeq 7\times 10^{40}~\mathrm{erg~s^{-1}}$, and $\mathrm{FWHM}(\Hbeta)\simeq 1200~\mathrm{km~s^{-1}}$ \citep{blanchard2017ps16dtm}. These properties result in $R_\mathrm{BLR}\simeq 0.009~\mathrm{pc}$, $M_\mathrm{BLR}\simeq 0.6~\Msun$, $\varepsilon\simeq 7\times 10^{-3}$, and $M_\mathrm{BH}\simeq 3\times 10^6~\Msun$. Because of the small BH mass, the AGN continuum could be affected by stellar light and $M_\mathrm{BH}$ may actually be slightly lower ($\simeq 10^{6}~\Msun$, e.g., \citealt{xiao2011lowmassagnbhmass}). Adopting the same way as in CSS100217, we obtain $L_\mathrm{bol}\simeq 9\times 10^{43}~\mathrm{erg~s^{-1}}$ and $\Gamma\simeq 0.3$. 

The $g$-band peak magnitude of PS16dtm is $-22$~mag \citep{blanchard2017ps16dtm}. Because it has not been faded away, total radiated energy in PS16dtm exceeds $\simeq 3\times 10^{51}~\mathrm{erg}$. To acquire this amount of radiation energy, $f_c M^\mathrm{BH}_\mathrm{ej}$ needs to be larger than 0.3~\Msun. If we assume that $f_c M^\mathrm{BH}_\mathrm{ej}$ is as much as $M_\mathrm{BLR}\simeq 0.6~\Msun$, $t_\mathrm{em}$ can be as long as $t_\mathrm{em}\simeq 110~\mathrm{days}$. PS16dtm is likely to start declining soon if it is powered by the BH disk wind as its duration matches $t_\mathrm{em}\simeq 110~\mathrm{days}$ now. The observed broad ($\sim 10,000~\kmps$) Mg~II absorption could be related to the fast BH disk wind.

\citet{blanchard2017ps16dtm} argue that PS16dtm is not due to an AGN variability because (i) they have luminosity increase much larger than those observed in other AGNs and (ii) viscous timescales in disks are much longer than timescales observed in STACs. However, both difficulties are overcome if we consider the interaction between the BH disk wind and the BLR clouds. An important characteristics of PS16dtm is the X-ray fading during the optical brightening. \citet{blanchard2017ps16dtm} suggest that TDE debris surrounding the X-ray emitting region obscures the X-ray emission from the center. As discussed in Section~\ref{sec:generalpicture}, the BH disk wind can obscure the central region and the X-ray fading is naturally expected in our BH disk wind model. In the case of PS16dtm, assuming that the ejecta mass is $\sim 0.5~\Msun$ and the inner and outer radii of the ejecta are $\sim 10^{13}$~cm and $10^{16}$~cm (a typical outer shock radius at around 100~days), respectively, the effective optical depth in the BH disk wind become $\tau_\mathrm{eff}\sim 3$. Therefore, the BH disk wind can actually obscure the central X-ray source. Because $\tau_\mathrm{eff}$ is close to unity, the degree of the obscuration in each STACs can easily change depending on, e.g., the mass and asphericity of the BH disk wind.

\section{Discussion}\label{sec:discussion}
\subsection{Mass ejection from BH accretion disks}\label{sec:bhmassejection}
A possible physical mechanism to initiate the 
transient disk winds from the BH accretion disks
is limit-cycle oscillations 
\citep[e.g.,][]{honma1991acc,szuszkiewicz1997acc,watarai2003acc}.
When the mass accretion rate of the disk is near-Eddington rate, the disk cycles between the gas-pressure dominated standard disk state and the slim disk state via the disk instability.
When it transforms from a gas-pressure dominated state
to a slim disk state, 
its luminosity can increase significantly 
and the luminosity can exceed the Eddington luminosity for a while. 
During this time, 
strong radiation-driven disk winds can be launched. 
By the two-dimensional radiation hydrodynamics simulations,
\cite{ohsuga2006limitcycle} has successfully reproduced 
the limit-cycle behavior and the transient radiation-driven winds,
of which the kinetic energy flux is above the Eddington luminosity
\citep[see also][]{teresi2004acc1,teresi2004acc2,ohsuga2007diskoutflow}.

The most famous candidate exhibits this kind of
limit-cycle oscillation is a microquasar GRS 1915+105.
The luminosity of GRS 1915+105
is comparable to the Eddington luminosity,
and this object clearly shows the quasi-periodic luminosity variation
with the timescale of several 10~sec \citep{belloni1997grs1915a,belloni1997grs1915b}.
Interestingly, the possible double-peak shape in the LC of PS16dtm is sometimes found in the microquasar GRS1915+105 \citep[e.g.,][]{janiuk2005grs1915}.
\cite{janiuk2011cyclebh} have reported that
the disk instability in the radiation pressure-dominated region
occurs in several stellar-mass accreting BHs ($M_\mathrm{BH}\sim 10~\Msun$)
with $\Gamma\gtrsim 0.1$.
Indeed, the AGNs hosting CSS100217 and PS16dtm have the Eddington ratio
of around 0.1 and are likely to
be able to trigger the limit-cycle oscillation\footnote{
The BH mass estimates by \citet{blanchard2017ps16dtm} are a factor of a few smaller than our estimates. This shows that the estimated Eddington ratios have uncertainties of a factor of a few. However, the exact Eddington ratio to trigger the limit-cycle oscillations is also uncertain \citep{janiuk2011cyclebh} and the uncertainties of a factor of a few are not critical in our model. The Eddington ratios in both estimates become near unity during the bursts.
}.
The timescales for stellar mass BH disks to have super-Eddington accretion
are $\sim 10-100~\mathrm{sec}$.
As the timescale is expected to be roughly proportional to
$M_\mathrm{BH}$,
the timescales for the BH disk wind from
$M_\mathrm{BH}\sim 10^6-10^8~\Msun$ are thought
to be $\sim 10-1000$ days.
These are comparable to or less than the emission timescale $t_\mathrm{em}$ (Eq.~\ref{eq:tem}). Thus, STACs usually radiate with the emission timescale in our model.

If the BH disk wind caused by the limit-cycle oscillation is responsible for the STAC activities,
STACs are expected to be brightened intermittently.
The prediction for the duration of the quiescent phase strongly depends
on the viscosity adopted in the disk model \citep{watarai2003acc}.
The quiescent phase can be more than 10 times longer
than the luminous phase
if the disk viscosity is sensitive to the radiation pressure.
If we simply assume that the properties of the viscosity remain unchanged, 
the duration of the quiescent phases is roughly proportional to the viscous timescale and therefore the central BH mass.
In the case of $M_\mathrm{BH}\sim 10^6-10^8~\Msun$,
the quiescent phase can last for several months to decades.

\begin{figure}
 \begin{center}
  \includegraphics[width=\columnwidth]{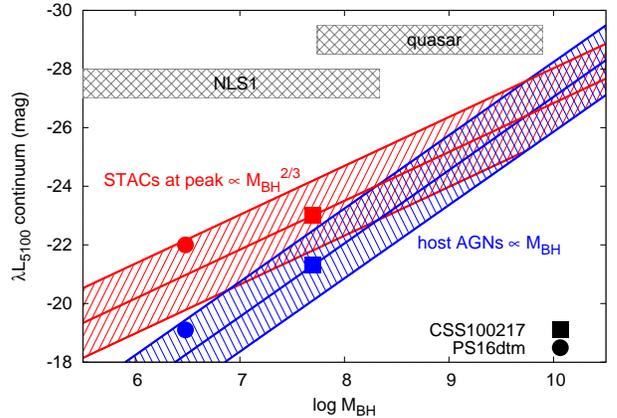}  
 \end{center}
\caption{
Luminosity estimates of STACs by the BH disk wind model and their host AGNs. Both luminosities are scaled with CSS100217 properties indicated with squares. We shade regions within a factor of 3 to account for uncertainties. The properties of PS16dtm shown with filled circles match the scaling expected from the BH disk wind model.
We also show the mass range of the central BHs in NLS1 galaxies \citep{zhou2006nls1} and that in
quasars \citep{shen2011agnbolcorr} which is obtained by removing 1\% outliers at the high and low mass ends.
}\label{fig:luminosity}
\end{figure}

\subsection{Diversity of STACs by the BH disk wind model}\label{sec:diversity}
We have proposed that STACs like CSS100217 and PS16dtm are related to AGN activities. We argue that STACs are preferentially observed in NLS1 galaxies because of their relatively small BH masses ($\sim 10^6-10^8~\Msun$, \citealt{zhou2006nls1}). Here, we consider diversities of STACs due to the diversities in BHs in AGNs in the context of our BH disk wind model.

The BLR density does not vary much due to their similar electron densities ($n_e\sim 10^9~\mathrm{cm^{-3}}$) and filling factors ($\varepsilon\sim 10^{-3}$, Eq.~\ref{eq:fillingfactor}). Also, the BH disk wind velocity is presumed to be similar ($v_\mathrm{ej}^\mathrm{BH}\simeq 0.1 c$). Therefore, the diversities in STACs in our BH disk wind model mainly come from the diversities in the total amount of the mass ejected from the central BH accretion disks. 

$M_\mathrm{ej}^\mathrm{BH}\simeq 4~\Msun$ is required to explain the observational properties of CSS100217.
Because $M_\mathrm{ej}^\mathrm{BH}$ can be proportional to $L_\mathrm{Edd}$ and $L_\mathrm{Edd}\propto M_\mathrm{BH}$, $M_\mathrm{ej}^\mathrm{BH}$ can be proportional to $M_\mathrm{BH}$. Then, the STAC luminosity $(E_\mathrm{kin}^\mathrm{BH}/t_\mathrm{em})$ is proportional to $M_\mathrm{BH}^{2/3}$ as $t_\mathrm{em}\propto M_\mathrm{ej}^\mathrm{BH\ 1/3}$. On the other hand, the AGN luminosities hosting STACs are presumed to be proportional to $M_\mathrm{BH}$ because $\Gamma\simeq 0.1$ is required to activate the limit-cycle oscillations. Therefore, when a BH mass increases by a factor of 10, a STAC luminosity increases only by a factor of 5 but an AGN luminosity increases by a factor of 10. 
In Fig.~\ref{fig:luminosity}, we show expected STAC and host AGN luminosity evolutions scaled with the properties of CSS100217.
STACs by the BH disk wind start to be ``buried'' in AGNs with the increasing AGN BH mass. On the contrary, they become easier to detect in low-mass AGNs, even though the STAC luminosities are also expected to decrease with smaller BH masses.
PS16dtm is observed in an AGN whose host $g$-band magnitude is $-19.1~\mathrm{mag}$ and its peak $g$-band magnitude is $-22$~mag. These properties roughly matches the above scaling based on CSS100217.

\subsection{Post-STAC luminosity evolution of CSS100217}\label{sec:css100217lumdecline}
CSS100217 has been monitored even after the major burst observed in 2010 and its light curve is available in the Catalina Real-Time Transient Survey website\footnote{\url{http://nesssi.cacr.caltech.edu/catalina/20100217/1002171400444123118p.html}}. Interestingly, the post-burst brightness of the CSS100217 host is about 0.5~mag fainter than the pre-burst brightness. The brightness difference in the host is likely associated with changes in the AGN accretion disk. If CSS100217 was a phenomenon that is not related to the central AGN activities, we do not naturally expect changes in the accretion disk shortly after the burst. Thus, the observed luminosity difference in the AGN hosting CSS100217 suggests that the transient is related to the AGN activity. In addition, the post-burst brightness is gradually recovering to the pre-burst brightness which matches a prediction of the limit-cycle oscillation \citep{watarai2003acc}. If the limit-cycle oscillation is actually triggering STACs, CSS100217 may become bright again in coming years or decades.

%\section{Conclusions}\label{sec:conclusions}
%We propose that superluminous transients from AGN central regions, or STACs, can originate from the interaction between a BH disk wind and BLR clouds. STACs currently known emit $\sim 10^{52}~\mathrm{erg}$ in a few hundred days \citep{drake2011css,blanchard2017ps16dtm}. If $\sim 1~\Msun$ with $\simeq 0.1c$ is ejected by disk instabilities such as limit-cycle oscillations that can occur near the Eddington luminosity, ejecta with kinetic energy of $\sim 10^{52}~\mathrm{erg}$ emerge from the vicinity of the central BH. If the BLR contains enough mass ($\gtrsim 1~\Msun$) to decelerate the ejecta, the ejecta kinetic energy can be efficiently converted to radiation in $\sim 100~\mathrm{days}$ and a superluminous transient with $\sim 10^{44}~\mathrm{erg~s^{-1}}$ appears at the AGN central region. We show that CSS100217 and PS16dtm can be explained by our model. If they are STACs triggered by the limit-cycle oscillations, they are likely to become bright again in coming years or decades.

\acknowledgments
This research is supported by the Grants-in-Aid for Scientific Research of the Japan Society for the Promotion of Science (TJM 16H07413, 17H02864; MT 15H02075, 15H00788; TM 16H02158, 16H01088; KO 15K05036).

\bibliographystyle{yahapj}
\bibliography{main}
%\begin{thebibliography}{}
%\end{thebibliography}

\end{document}